\documentclass[twocolumn,showpacs,preprintnumbers,amsmath,amssymb,pra]{revtex4}
\usepackage{graphicx}% Include figure files
\usepackage{dcolumn}% Align table columns on decimal point
\usepackage{bm}% bold math

\begin{document}

\preprint{APS/123-QED}

\title{Internal Josephson Effects in Spinor Dipolar Bose--Einstein Condensates}% Force line breaks with \\

\author{Masashi Yasunaga}
% \altaffiliation[Also at ]{Department of Physics, Osaka City University}%Lines break automatically or can be forced with \\
\author{Makoto Tsubota}%
 \email{tsubota@sci.osaka-cu.ac.jp}
\affiliation{
Department of Physics, Osaka City University, Sumiyoshi-ku, Osaka 558-8585, Japan
}%

\date{\today}% It is always \today, today,

\begin{abstract}

We theoretically study the internal Josephson effect, which is driven by spin exchange interactions and   magnetic dipole-dipole interactions, in a three-level system for spin-1 Bose--Einstein condensates, obtaining novel spin dynamics. We introduce single spatial mode approximations into the Gross--Pitaevskii equations and derive the Josephson type equations, which are analogous to tunneling currents through three junctions between three superconductors. From an analogy with two interacting nonrigid pendulums, we identify unique varied oscillational modes, called the $0$--$\pi$, $0$--$running$, $running$--$running$, $2n\pi\ \& \ running$--$2\pi$, $single \  nonrigid \  pendulum$, and $two \  rigid \  pendulums$ phase modes. These Josephson modes in the three states are expected to be found in real atomic Bose gas systems. 
	 
\end{abstract}

\maketitle

\section{INTRODUCTION}\label{sec:in}

The Josephson effect is a universal quantum phenomenon, defined as a current between two or more macroscopic quantum states with weak coupling driven by the relative phases of the macroscopic wave functions. The effect for a junction of two superconductors was originally predicted by Josephson \cite{Josephson1962} and discovered by Anderson and Rowell \cite{Anderson1963}. When the effect was first predicted and discovered it was considered a characteristic phenomenon of superconductors.  However, the effect has been found in diverse fields since Feynman redefined the Josephson effect as the quantum tunneling between two levels \cite{Feynman1965}. In particular, Maki and Tsuneto have considered nonlinear ringing between equal spin pairing states $|\uparrow\uparrow\rangle$ and $|\downarrow\downarrow\rangle$ in superfluid $^3$He-A as a Josephson junction between the internal degrees of freedom \cite{Maki1974}. Webb {\it et al.} subsequently observed the effect by NMR \cite{Webb1974}. 

Such Josephson effects have also been studied in atomic Bose--Einstein condensates (BECs). Smerzi {\it et al.} derived Josephson type equations in a double well potential from Gross--Pitaevskii (GP) equations with two-mode approximations, revealing three oscillations, namely the $0$, $\pi$ and {\it running} phase modes \cite{Smerzi1997,Raghavan1999}. Macroscopic quantum self trapping from the $running$ mode and AC Josephson effects from the $0$ phase mode have been observed by Albiez {\it et al.} \cite{Albiez2005}. Furthermore, Zhang {\it et al.} considered transitions between three energy states for spin-1 BECs driven by spin exchange interactions as internal Josephson effects\cite{Zhang2005} and Chang {\it et al.} observed the internal Josephson oscillation \cite{Chang2005}.

\begin{figure}[b]
\begin{center}
\includegraphics[width=0.8\linewidth]{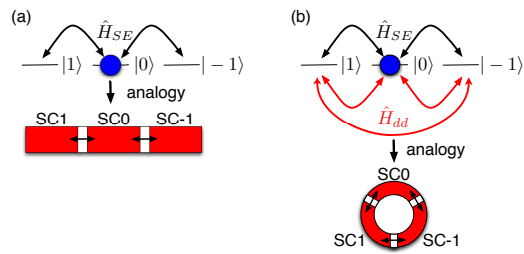} 
\caption{ (color online) The transitions, driven by the Hamiltonians of a spin exchange interaction $\hat{H}_{SE}$ (black arrows) and an MDDI $\hat{H}_{dd}$ (red arrows), between the energy states (solid line) are analogous to the Josephson effect for cooper pairs tunneling through (a) two and (b) three insulators (white boxes) between super conductors, named SC1, SC0, and SC-1 (red boxes). (a) and (b) compare differently with the MDDI, was the MDDI can also drives transition from $|\pm 1\rangle$ to $|\mp 1\rangle$. The blue circles represent the condensate. }
\label{fig:analogy12}
\end{center}
\end{figure}

In addition, studies have focused on the magnetic dipole--dipole interaction (MDDI) in BECs. The interaction between spins has a characteristic symmetry for spin and orbit. Therefore, the interaction is theoretically expected to give new quantum phases \cite{Yi2004, Yuki2006a, Makela2007}, the Einstein--de Haas effect \cite{Yuki2006b} and so on. Griesmaier {\it et al.} realized spinor dipolar condensates using $^{52}$Cr atoms \cite{Griesmaier2005} which have larger magnetic moments than alkali atoms. The condensates clearly show shape-dependent anisotropy of the MDDI \cite{Stuhler2006, Lahaye2007}. Thus, the study of the MDDI has opened new avenues for investigating spinor condensates.

For the MDDI, we can demonstrate new internal Josephson effects in spin-1 BECs. The internal Josephson effect without the MDDI is analogous with two junctions between three superconductors because the weak coupling, namely the spin exchange interaction, permits transitions only from $|0\rangle$ to $|\pm 1\rangle$ and from $|\pm 1\rangle$ to $|0\rangle$. However, it is possible to transfer from $|\pm 1\rangle$ to $|\mp 1\rangle$ if the MDDI drives the transitions because the anisotropic interaction breaks the conservation law of magnetization, as will be discussed later. As the previous study, Cheng {\it et. al. } have discussed the Josephson effects driven by the MDDI in the conservation law \cite{Cheng2005}. Therefore, we consider out of the law that the transitions through the MDDI are analogous with Josephson effects for three circular junctions between three superconductors (see Fig. \ref{fig:analogy12}).

In this paper, we introduce the spin dynamics in spinor dipolar BECs in Sec. \ref{sec:formulation} using the GP equations with single mode approximations, and we demonstrate new Josephson oscillations in Sec. \ref{sec:dynamics}. Section \ref{sec:con} is devoted to a conclusion.
	  
\section{FORMULATION}\label{sec:formulation}

\subsection{Single mode approximation}

Before we consider the internal Josephson effect for spin-1 BECs, we analyze the spin-1 GP equations with an MDDI:
\begin{eqnarray}\label{eq:GP}
i\hbar\frac{\partial \psi_\alpha}{\partial t} &=& \left(-\frac{\hbar^2}{2M}\nabla^2+V-\mu\right)\psi_\alpha\nonumber \\
&& - g_e\mu_B {\bf H}\cdot {\bf F}_{\alpha \beta}\psi _\beta \nonumber\\
&&+ c_0\psi _\beta^*\psi _\beta\psi _\alpha + c_2 {\bf F}\cdot{\bf F}_{\alpha \beta}\psi _\beta\nonumber \\
&&+c_{dd}\int d{\bf r}' \frac{\delta_{ij}-3e^ie^j}{|{\bf r}-{\bf r}'|^3}F_i({\bf r}')F_{\alpha \beta}^j\psi_\beta, 
\end{eqnarray}
where the integers $\alpha$ and $\beta$ represent the spin $0 and \pm 1$ states, $V({\bf r})$ is the trapping potential, and $\mu$ the chemical potential. The spin density vector ${\bf F} =(F_x, F_y, F_z)$ and the vector of the spin components ${\bf F}_{\alpha\beta} = (F_{\alpha\beta}^x, F_{\alpha\beta}^y, F_{\alpha\beta}^z)$ are represented by $F_{i = x,y,z} = \psi_\alpha^*F_{\alpha \beta}^i\psi_\beta$ given by the components $F_{\alpha \beta}^i$ of the spin matrices $\hat{F}_i$ for spin-1. The Zeeman term is presented by Lande's $g$-factor of an electron $g_e$, Bohr magneton $\mu _B$, and external magnetic field ${\bf H}$. The coefficients $c_0 = (g_0+2g_2)/3$ and $c_2 = (g_2-g_0)/3$ are the interaction parameters for $g_i = 4\pi\hbar^2a_i/M$ represented by the s-wave scattering lengths $a_i$. The coefficient of the MDDI is given by $c_{dd} = \mu_0g_e^2\mu_B^2/4\pi$ with ${\bf e} = ({\bf r}-{\bf r}')/|{\bf r}-{\bf r}'|$ being a unit vector. 

Since it is difficult to derive the internal Josephson effect from Eq. (\ref{eq:GP}) directly, we introduce the single mode approximation \cite{Zhang2005}:
\begin{equation}\label{eq:SMA}
\psi_i({\bf r},t) = \sqrt{N}\xi_i(t)\phi({\bf r})\exp\left(-\frac{i\mu t}{\hbar}\right),
\end{equation}
where $\phi$ satisfies the eigenvalue equation $(-\hbar^2\nabla ^2/2M+V+c_0n)\phi = \mu \phi$ with the relation $\int d{\bf r} |\phi|^2 = 1$. The approximation can be used when the shapes of the condensates are decided by the spin independent terms, namely $|c_0| \gg |c_2|$. The condition is satisfied for $^{87}$Rb and $^{23}$Na. The validity of the approximation in spin-1 dipolar BECs has been discussed by Yi and Pu \cite{Yi2006}. They found that the approximation is affected by the anisotropic parameter of the shape of the condensates $\lambda$ and the ratio $c_{dd}/|c_2|$. For small $\lambda$, the approximation is well defined even if the ratio is large. Here, we assume that the ratio is smaller than $1$, allowing the approximation to be used. Hence, introducing Eq. (\ref{eq:SMA}) into Eq. (\ref{eq:GP}), we can derive the equation of the spinor $|\xi \rangle = (\xi _1, \xi _0, \xi _{-1})^T$ in ${\bf H} = H\hat{z}$,
\begin{equation}\label{eq:GPSMA1}
i\hbar\frac{d}{dt}|\xi\rangle = (\hat{H}_Z+\hat{H}_{SE}+\hat{H}_{dd})|\xi \rangle,
\end{equation}
given by the Zeeman, the spin exchange interaction, and the MDDI Hamiltionians,
\begin{eqnarray}
\hat{H}_Z &=& -g\mu _B\left(
\begin{array}{ccc}
H&0&0\\
0&0&0\\
0&0&-H
\end{array}
\right) \label{eq:HZ}\\
\hat{H}_{SE} &=& c\left(
\begin{array}{ccc}
1-2\rho _{-1}&\rho _{-10}&0\\
\rho _{0-1}&1-\rho _0&\rho _{01}\\
0&\rho _{10}&1-2\rho _1
\end{array}
\right)\label{eq:Hint}\\
\hat{H}_{dd} &=& c_{dd}^+\left(
\begin{array}{ccc}
\rho _0&\rho _{-10}&0\\
\rho _{0-1}&1-\rho _{0}&\rho _{01}\\
0&\rho _{10}&\rho _{0}
\end{array}
\right)\nonumber\\
&&+c_{dd}^-\left(
\begin{array}{ccc}
0&\rho _{10}&\rho _0\\
\rho _{01}&\rho _{-11}+\rho _{1-1}&\rho _{0-1}\\
\rho _0&\rho _{-10}&0
\end{array}
\right)\nonumber\\
&&+2c_{dd}^z\left(
\begin{array}{ccc}
m&0&0\\
0&0&0\\
0&0&-m
\end{array}
\right),\label{eq:Hdd}
\end{eqnarray}
where $\rho _{ij} = \xi _i^*\xi _j$ ($\rho _{ii} = \rho _i$) are components of the density matrix $\rho = |\xi \rangle\langle \xi |$, $c_{dd}^\pm = c_{dd}^x \pm c_{dd}^y$ with $c_{dd}^z$ given by
\begin{equation}\label{eq:cdd}
c_{dd}^i =\frac{c_{dd}}{2}N\int\int d{\bf r}d{\bf r}' \frac{|\phi ({\bf r})|^2|\phi ({\bf r}')|^2}{|{\bf r}-{\bf r}'|^3}(1-3e^i\sum _j e^j),
\end{equation}
$c = c_2N\int d{\bf r}|\phi|^4$ are the interaction parameters, and $m = \rho _1 -\rho _{-1}$ is the magnetization. Comparing Eq. (\ref{eq:Hint}) with Eq. (\ref{eq:Hdd}), we can conclude that the term with $c_{dd}^-$ in $\hat{H}_{dd}$ includes the operators $| \pm 1 \rangle \langle \mp 1|$, which project from $|\mp 1\rangle$ to $|\pm 1\rangle$, where $|1\rangle = (1, 0, 0)^T$, $|-1\rangle = (0, 0, 1)^T$ are unit vectors. These operators are not included in $\hat{H}_{SE}$. We again emphasize that the transition from $|\pm 1\rangle$ to $|\mp 1\rangle$ is possible only by the MDDI.

Using the relation $\sum _i \rho _i = 1$ we can rewrite the spin exchange and the dipole term to give
\begin{eqnarray}
\hat{H}_{SE}+\hat{H}_{dd} &=& (c+c_{dd}^+)\left(
\begin{array}{ccc}
\rho _{0}&\rho _{-10}&0\\
\rho _{0-1}&1-\rho _0&\rho _{01}\\
0&\rho _{10}&\rho _0
\end{array}
\right)\nonumber\\
&&+c_{dd}^-\left(
\begin{array}{ccc}
0&\rho _{10}&\rho _0\\
\rho _{01}&\rho _{-11}+\rho _{1-1}&\rho _{0-1}\\
\rho _0&\rho _{-10}&0
\end{array}
\right)\nonumber\\
&&+(c+2c_{dd}^z)\left(
\begin{array}{ccc}
m&0&0\\
0&0&0\\
0&0&-m
\end{array}
\right).\label{eq:Hint+Hdd}
\end{eqnarray}
Equation (\ref{eq:Hint+Hdd}) clearly shows that a $c$ term can be included in the $c_{dd}^+$ and $c_{dd}^z$ terms. In short, the MDDI is partly renormalized in the spin exchange interaction. 

In conclusion for the MDDI, three important roles for spin dynamics appear in Eq. (\ref{eq:Hdd}). The first is the spin exchange effect, as in Eq. (\ref{eq:Hint}), given by the $c_{dd}^+$ and $c_{dd}^z$ terms of $\hat{H}_{dd}$. The second is the transition from the $\pm 1$ to the $\mp 1$ states, given by the $c_{dd}^-$ term. The last is an interaction between the magnetization and the spin, given by the $c_{dd}^z$ term, which implies that the magnetization produces a molecular field. 

\subsection{Josephson type equations}\label{ss:jte}

In order to investigate the Josephson effect for transitions between the three states, we substitute $\xi _j = \sqrt{\rho _j}e^{i\theta _j}$ in Eq. (\ref{eq:GPSMA1}), deriving the Josephson type equations for $\rho _0$, $m$ and relative phases $\theta = \theta _1 + \theta _{-1}-2\theta _0$ and $\theta _m = \theta _{1}-\theta _{-1}$;
\begin{eqnarray}
\dot{\rho _{0}} &=& \frac{2}{\hbar}c'\rho _0\sqrt{(1-\rho _0)^2-m^2}\sin\theta \nonumber\\
&+&\frac{2}{\hbar}c_{dd}^-\rho _0\{(1-\rho _0)\sin\theta\cos\theta _m+m\cos\theta\sin\theta _m\}, \label{eq:JE1} \\ \nonumber\\
\dot{m}&=&-\frac{2}{\hbar}c_{dd}^-\rho _0\{\sqrt{(1-\rho _0)^2-m^2}\sin\theta _m \nonumber\\
&+& (1-\rho _0)\cos\theta\sin\theta _m+m\sin\theta\cos\theta _m\}, \label{eq:JE2}\\\nonumber\\
\dot{\theta} &=& \frac{2}{\hbar}c'(1-2\rho _0)+\frac{2}{\hbar}c'\frac{(1-\rho _0)(1-2\rho _0)-m^2}{\sqrt{(1-\rho _0)^2-m^2}}\cos\theta \nonumber\\
&+&\frac{2}{\hbar}c_{dd}^-\Biggl\{\frac{(1-\rho _0)(1-2\rho _0)-m^2}{\sqrt{(1-\rho _0)^2-m^2}}\cos\theta _m\nonumber\\
&+&(1-2\rho _0)\cos\theta\cos\theta _m-m\sin\theta\sin\theta _m\Biggr\}, \label{eq:JE3} \\ \nonumber\\
\dot{\theta _m} &=& -\frac{2}{\hbar}(E_Z+c''m)+\frac{2}{\hbar}c'\frac{m\rho _0}{\sqrt{(1-\rho _0)^2-m^2}}\cos\theta \nonumber \\
&+&\frac{2}{\hbar}c_{dd}^-\left\{\frac{m\rho _0}{\sqrt{(1-\rho _0)^2-m^2}}\cos\theta _m+\rho _0\sin\theta\sin\theta _m\right\}, \nonumber\\\label{eq:JE4}
\end{eqnarray}		  
where $E_Z = -g_e\mu _B H$ is the Zeeman energy and $c' = c+c_{dd}^+$ and $c''=c+2c_{dd}^z$ are the interaction coefficients. The equations clearly demonstrate the definition of the Josephson effect, namely a current between states driven by a relative phase.

\begin{figure}[t]
\begin{center}
\includegraphics[width=0.9\linewidth]{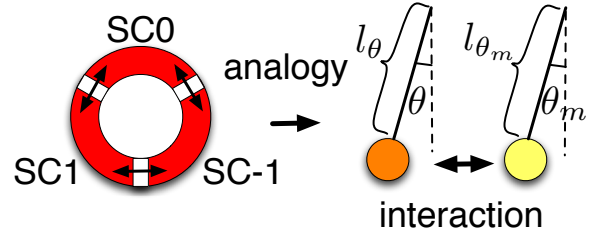} 
\caption{(color online) Analogy between the dynamics of Josephson junctions and two interacting nonrigid pendulums with lengths $l_\theta \propto c'\rho _0\sqrt{(1-\rho _0)^2-m^2}$ and $l _{\theta _m} \propto (c_{dd}^-/c')l_\theta$.}
\label{fig:analogy3}
\end{center}
\end{figure}

Firstly, from Eq. (\ref{eq:JE2}) we can easily see that conservation of the magnetization is broken by the $c_{dd}^-$ term. When $c_{dd}^- = 0$, the magnetization becomes a conserved value. Under the single mode approximation, $c_{dd}^i=0$ for uniform density and a spherical shape, which is obtained from the integration, Eq. (\ref{eq:cdd}). Secondly, we note that the variables $\rho _0$ and $m$ are canonical conjugates of $\theta$ and $\theta _m$, respectively. Therefore, these equations can be derived from the canonical equations of motions:
\begin{equation}
\dot{\rho _0} = -\frac{2}{\hbar}\frac{\partial {\cal H}}{\partial \theta}, \  \dot{\theta} = \frac{2}{\hbar}\frac{\partial {\cal H}}{\partial \rho _0}, \  \dot{m} = \frac{2}{\hbar}\frac{\partial {\cal H}}{\partial \theta _m}, \ {\rm and} \ \dot{\theta _m} = -\frac{2}{\hbar}\frac{\partial {\cal H}}{\partial m}, \nonumber
\end{equation} 
whose Hamiltonian is represented as
\begin{equation}\label{eq:H}
{\cal H}={\cal H}_\theta+{\cal H}_{\theta _m}+{\cal H}_{int},
\end{equation}
where
\begin{eqnarray}
{\cal H}_\theta&=& c'\{\rho _0(1-\rho _0)+\rho _0\sqrt{(1-\rho _0)^2-m^2}\cos\theta\},\label{eq:H_theta}\\
{\cal H}_{\theta _m}&=&\frac{c''}{2}m^2+E_zm+c_{dd}^-\rho _0\sqrt{(1-\rho _0)^2-m^2}\cos\theta _m, \nonumber\\\label{eq:H_theta_m}\\
{\cal H}_{int}&=&c_{dd}^-\{\rho _0(1-\rho _0)\cos\theta\cos\theta _m-\rho _0m\sin\theta\sin\theta _m\}.\nonumber\\\label{eq:H_int}
\end{eqnarray}
Since the Hamiltonian ${\cal H}(\rho _0 (t), m(t), \theta (t), \theta _m (t))$ is not explicitly time dependent, the relation $d{\cal H}/dt = \partial {\cal H}/\partial t = 0$ is satisfied. A comparison of the Hamiltonian with that of a pendulum of length $l$ and angle $\varphi$, ${\cal H}_p(P, \varphi) = P^2/2M + Mgl\cos \varphi$, shows that the dynamics described by the equations may be equal to that of two nonrigid pendulums, because ${\cal H}_\theta$ and ${\cal H} _{\theta _m}$ are similar to the Hamiltonian of nonrigid pendulums with lengths proportional to $c'\rho _0\sqrt{(1-\rho _0)^2-m^2}$ and $c_{dd}^-\rho _0\sqrt{(1-\rho _0)^2-m^2}$ and angles $\theta$ and $\theta _m$, respectively. The two pendulums have some interaction with each other, represented by terms in ${\cal H}_{int}$ that include the product of $\theta$ by $\theta _m$. Therefore, we consider that Eq. (\ref{eq:H}) represents the Hamiltonian of two interacting nonrigid pendulums (see Fig. \ref{fig:analogy3}).

\section{SPIN DYNAMICS}\label{sec:dynamics}

In this section, we discuss solutions of the equations for $E_Z = 0$, namely small oscillations of two nonrigid pendulums in Sec. \ref{ssec:small}, oscillations of a single nonrigid pendulum in Sec. \ref{ssec:snr} and oscillations of two interacting nonrigid pendulums in Sec. \ref{ssec:dnr}. 

\subsection{Small oscillation around stationary solutions}\label{ssec:small}

Not being able to solve the Josephson type equation directly, we first obtain the solutions for the stationary states:
\begin{equation}\label{eq:ss1}
\rho _0 = \rho _a, \hspace{5 mm} m = 0,  \hspace{5 mm} \theta = (2n+1)\pi, \hspace{5 mm} \theta _m = \theta _a
\end{equation}  
and
\begin{equation}\label{eq:ss2}
\rho _0 = \frac{1}{2}, \hspace{5 mm} m = 0,  \hspace{5 mm} \theta = 2n\pi, \hspace{5 mm} \theta _m = n\pi.
\end{equation}
 Here $\rho _a$ and $\theta _a$ are arbitrary constants. For $E_Z = 0$, we cannot obtain the stationary magnetization $m \neq 0$. In order to study the features of the oscillation of the pendulums, we perform linear analysis by introducing the approximations $\rho _0 \simeq \rho _{0s} + \delta \rho _0$, $m \simeq m_s + \delta m$, $\theta \simeq \theta _s + \delta \theta$ and $\theta _m \simeq \theta _{ms}+\delta \theta _m$, where these deviations from the stationary solutions are very small, into the Josephson type equations.  Linearizing the equations by ignoring squares of the deviations, we obtain the equations for the deviations; the solutions around Eq. (\ref{eq:ss1}) are
\begin{equation}\label{eq:tss1}
\dot{\delta m} = \dot{\delta \theta} = 0
\end{equation}
and those around Eq. (\ref{eq:ss2}) are 
\begin{eqnarray}\label{eq:ho}
\ddot{\delta \rho _0} &=& -\omega _{\rho _0\pm}^2\delta \rho _0 \label{eq:tss2_r}\\
\ddot{\delta m} &=& -\omega _{m\pm}^2\delta m \label{eq:tss2_m},
\end{eqnarray}
 where $\omega _{\rho _0\pm}^2 = 4(c'\pm c_{dd}^-)^2/\hbar^2$ and $\omega _{m\pm}^2 = 4c_{dd}^-\{\mp c_{dd}^z\pm(c_{dd}^+\pm c_{dd}^-)/2\}/\hbar^2$ with $\pm$ indicating the equations for $\theta _{ms}= 2n\pi$ and $(2n+1)\pi$, respectively. From Eq. (\ref{eq:tss1}), we can conclude that there are no stable oscillational solutions around Eq. (\ref{eq:ss1}). On the other hand, Eqs. (\ref{eq:tss2_r}) and (\ref{eq:tss2_m}) are easily solved and have the solutions $\delta \rho _0 = \Delta _\rho \cos\omega _{\rho\pm}t$ and $\delta m = \Delta _m \cos\omega _{m\pm}t$, whose amplitudes are $\Delta _\rho$ and $\Delta _m$. Considering the deviations from Eqs. (\ref{eq:tss1}) to (\ref{eq:tss2_m}), we can conclude that there are no oscillations around $\theta = (2n+1)\pi$, namely the $\pi$ phase mode in the canonically conjugate variable $\{\rho _0, \theta\}$, whereas there are $0$ and $\pi$ phase modes in $\{m, \theta _m\}$.

\subsection{Josephson effect for a nonrigid pendulum}\label{ssec:snr}

\begin{figure}[t]
\begin{center}
\includegraphics[width=0.7\linewidth]{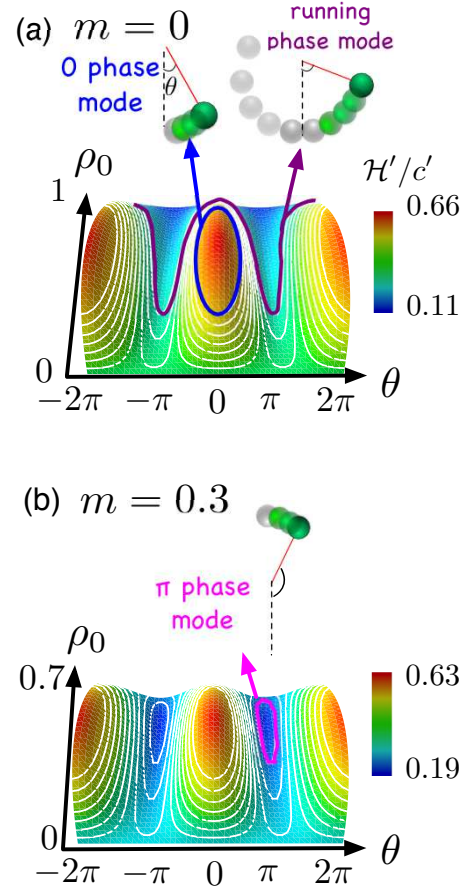} 
\caption{(Color online) Contour lines of Eq. (\ref{eq:Hcdd-0}) for (a) $m = 0$ and (b) $0.3$ with $\delta _{dd} = 0.3c'$ $(>0)$, represented as white. The blue, purple and magenta lines show the $0$, $running$ and $\pi$ phase modes, respectively. }
\label{fig:contour1}
\end{center}
\end{figure}

Next, in order to review the $0$, $\pi$, and $running$ phase modes in spin-1 BECs, we study the Josephson effect for $c_{dd}^- =0$ obtained for spherically shaped condensates. Under this condition, the spin dynamics are represented by Eqs. (\ref{eq:JE1}) and (\ref{eq:JE2}) with $c_{dd}^- = 0$, where there is no need to consider $\dot{m}$ and $\dot{\theta _m}$ because $m$ is constant. The equations are the same as for the Josephson effect driven by the spin exchange interaction in Ref. \cite{Zhang2005}  if the quadratic Zeeman effect, which is effective for the hyperfine interaction between the nuclear and electron spins in the Zeeman splits \cite{Stenger1998, Chang2005}, is introduced in the equation. Therefore, considering the quadratic Zeeman effect $\delta _{dd} \propto H_{eff}^2$ given by the effective dipole magnetic field $H_{eff} \propto c_{dd}^zm $, we obtain
\begin{eqnarray}
\dot{\rho _0} &=& \frac{2}{\hbar}c'\rho _0\sqrt{(1-\rho _0)^2-m^2}\sin\theta \label{eq:JE2_1}\\
\dot{\theta} &=& -\frac{2}{\hbar}\delta _{dd}+\frac{2}{\hbar}c'(1-2\rho _0)\nonumber \\
&&+\frac{2}{\hbar}c'\frac{(1-\rho _0)(1-2\rho _0)-m^2}{\sqrt{(1-\rho _0)^2-m^2}}\cos\theta, \label{eq:JE2_2}
\end{eqnarray}  
where $\rho _0$ is the canonical conjugate of $\theta$. Hence, these equations are derived from the canonical equations of motions for the Hamiltonian:
\begin{eqnarray}\label{eq:Hcdd-0}
{\cal H}' &=&\delta _{dd}(1-\rho _0)+c'\{\rho _0(1-\rho _0) \nonumber\\
&&+ \rho _0\sqrt{(1-\rho _0)^2-m^2}\cos\theta.
\end{eqnarray}
The Hamiltonian directly represents the system of a nonrigid pendulum. Using ${\cal H}' = {\cal H}'_0 = {\cal H}'(\rho _0 (0), \theta (0))$ for $\dot{{\cal H}'} = 0$, we can solve Eq. (\ref{eq:JE2_1}) and obtain the solutions expressed by Jacobi's elliptic function ${\rm cn}(a, k)$; for $ c_{dd}^+ > 0$
\begin{equation}
\rho _0 = \rho _b+(\rho _c-\rho _b){\rm cn}^2\left(\frac{1}{\hbar}\sqrt{2c'\delta _{dd}(\rho _c-\rho _a)}t, k\right),
\end{equation} 
and for $c_{dd}^+ < 0$
\begin{equation}
\rho _0 = \rho _b-(\rho _b-\rho _a){\rm cn}^2\left(\frac{1}{\hbar}\sqrt{-2c'\delta _{dd}(\rho _c-\rho _a)}t, k\right),
\end{equation}
where $\rho _{a, b, c}$ are the roots of $\dot{\rho _0} = 0$, and $k = \{(\rho _c -\rho _b)/(\rho _c -\rho _a)\}^{1/2}$. Pendulum-like oscillations occur on the contour lines of Eq. (\ref{eq:Hcdd-0}) for energy being conserved, shown in Fig. \ref{fig:contour1}. The dynamics can be classified into three modes; the dynamics on the lines around $\theta = 0$ and $\pi$ are called the $0$ and $\pi$ phase modes respectively and the line from $\theta = -2\pi$ to $2\pi$ is the {\it running} phase mode. The $0$ phase mode corresponds to the motion of a pendulum oscillating around $\theta = 0$ with a varying length. The $\pi$ phase mode also shows an oscillation around $\theta = \pi$,a characteristic dynamics for the nonrigid pendulum. Finally, the {\it running} phase mode represents the rotational dynamics.

\subsection{Josephson effect for two pendulums}\label{ssec:dnr}

Here we discuss the dynamics in Eqs. (\ref{eq:JE1}) to (\ref{eq:JE4}). Solving the equations numerically using the fourth-order Runge--Kutta method, we obtain several characteristic results for two interacting nonrigid pendulums. In the calculations, we estimate that the interaction parameters satisfy the relations $c'/c_{dd}^- = 11$ and $c''/c_{dd}^- = 12$, which are determined from the order estimations $c_{dd}^+ \sim c_{dd}^- \sim c_{dd}^z$ and $c \sim 10c_{dd}^+$. The orders between the spin exchange interactions and MDDIs have been discussed by Yi and Pu \cite{Yi2008}.  In this subsection, we denote the pendulums having angles $\theta$ and $\theta _m$ as $\theta$ and $\theta _m$ pendulums, respectively, considering the dynamics from the time developments of the canonically conjugate variables (in Sec. \ref{sssec:modes}) and the Hamiltonian (in Sec. \ref{sssec:hamiltonian}).

\subsubsection{Modes in the dynamics of four variables}\label{sssec:modes}

 \begin{figure}[t]
\begin{center}
\includegraphics[width=0.9\linewidth]{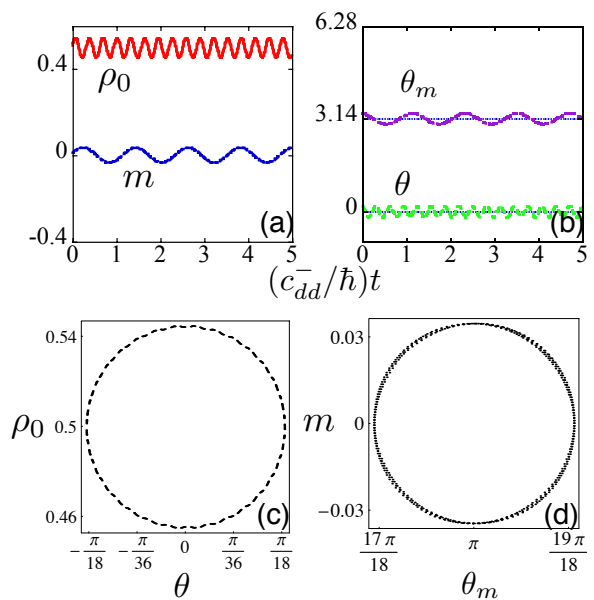} 
\caption{ (color online)  Typical $0$--$\pi$ phase mode. The dynamics of the four variables $\rho _0$, $m$, $\theta$ and $\theta _m$ in (a) and (b) and the phase projections of the motions in $\{\rho _0, \theta\}$ and $\{m, \theta _m\}$ in (c) and (d) are given for  $\rho _0 (0) = 0.51$, $m (0) = 0.01$, $\theta (0) = \pi/18$, $\theta _m (0) = 19\pi/18$.}
\label{fig:0-pi1}
\end{center}
\end{figure}

 \begin{figure}[t]
\begin{center}
\includegraphics[width=0.9\linewidth]{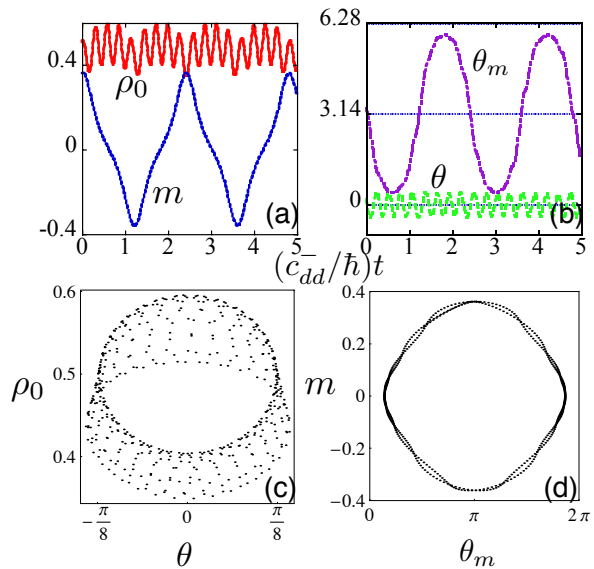} 
\caption{ (color online)  Quasi-$0$--$\pi$ phase mode for $\rho _0 (0) = 0.51$, $m (0) = 0.36$, $\theta (0) = \pi/18$, $\theta _m (0) = 19\pi/18$.}
\label{fig:0-pi2}
\end{center}
\end{figure}

 \begin{figure}[t]
\begin{center}
\includegraphics[width=0.95\linewidth]{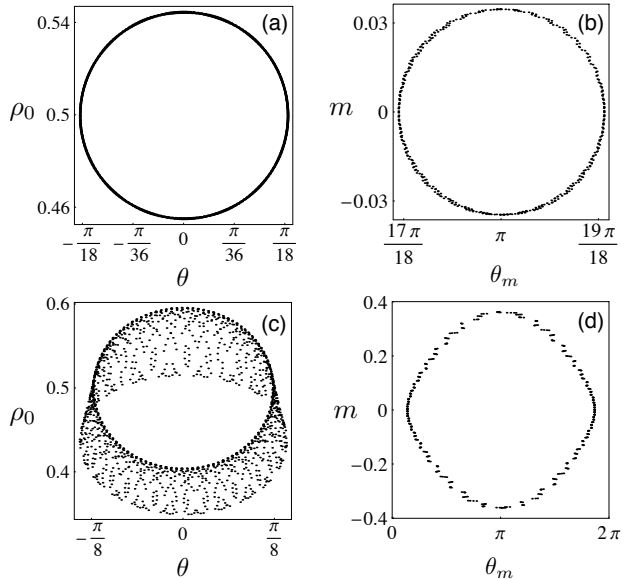} 
\caption{Poincar\'e mappings for the $0$--$\pi$ modes within $(c_{dd}^-/\hbar)t = 500$. The upper and lower figures show the projections of the Poincar\'e sections at $T = 2\pi/\omega _{\rho _0 -}$ for the phase spaces $\{\rho _0, \theta\}$ and $ 2\pi/\omega _{m-}$ for $\{m, \theta _m\}$ in Fig. \ref{fig:0-pi1} and Fig. \ref{fig:0-pi2}, respectively.}
\label{fig:Poincare}
\end{center}
\end{figure}
First, we present the typical $0$--$\pi$ phase mode for which $\theta$ and $\theta _m$ oscillate around $\theta = 0 $ and $\theta _m = \pi$, respectively; shown in Fig. \ref{fig:0-pi1} (a) and (b).   The phase mode is clearly obtained by applying small deviations from the stationary solution, Eq. (\ref{eq:ss2}). Considering the motion of the pendulums, the $\theta$ and $\theta _m$ pendulums make small oscillations around $0$ and $\pi$ as the lengths grow and shrink. The features of the oscillations are represented by trajectories projected in the phase spaces $\{ \rho _0, \theta\}$  Fig. \ref{fig:0-pi1} (c) and $\{m, \theta _m\}$ (d). Since the trajectory is confined to a simple closed orbit, we can see that the motion of the pendulums is similar to that of two independent pendulums, namely a periodic motion.  On the other hand, changing $m (0) = 0.01$ to $m (0) =0.36$, we obtain the quasi-$0$--$\pi$ phase mode, shown in Fig. \ref{fig:0-pi2}. The trajectory plotted in Fig. \ref{fig:0-pi2} (c) is not an orbit and hence we regard the spread trajectory as evidence of chaotic motion of the $\theta$ pendulum. In order to investigate the chaotic motions, we plot Poincar\'e mappings in Fig. \ref{fig:Poincare}. The mappings represent projections of the cross sections at $t = nT$ $(n = 1, 2, \ldots)$, where the period $T$ is given by eigenfrequencies in the systems, providing direct evidence of chaotic motion for the $0$--$\pi$ phase modes, because there is only one point in the Poincar\'e mappings where the motions are not chaotic but periodic. However, it is unexpected that the typical $0$--$\pi$ phase mode should be chaotic. Therefore, we conclude that the motions of the two pendulums are essentially chaotic and thus the dynamics do not occur on a contour line of the $0$ phase mode. The $\theta _m$ pendulum is also not a dynamic on a contour line of the $\pi$ phase mode.

 \begin{figure}[t]
\begin{center}
\includegraphics[width=0.9\linewidth]{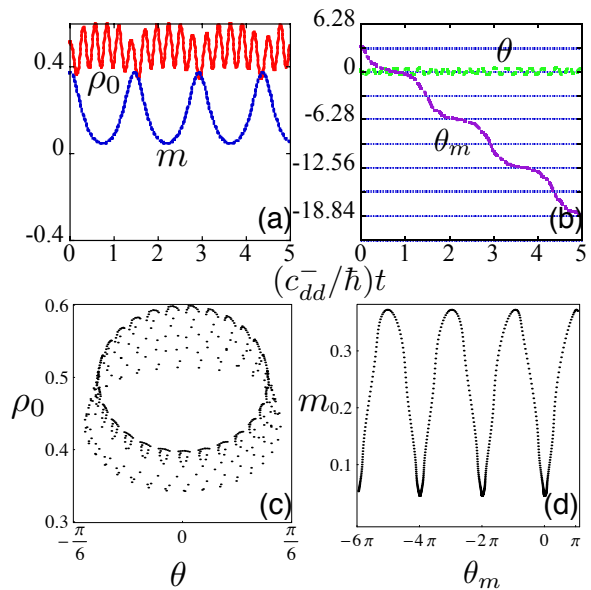} 
\caption{ (color online)  $0$--$running$ phase mode for $\rho _0 (0) = 0.51$, $m (0) = 0.37$, $\theta (0) = \pi/18$, $\theta _m (0) = 19\pi/18$.}
\label{fig:0-pi3}
\end{center}
\end{figure}

Setting $m (0) = 0.37$, where the other initial conditions are same as the $0$--$\pi$ phase mode, we can produce the $0$--$running$ phase mode in Fig. \ref{fig:0-pi3}. The phase mode shows that the $\theta$ pendulum oscillates around $\theta = 0$ with a small amplitude chaotically, whereas the $\theta _m$ pendulum rotates continuously, given by the dynamics of $\theta _m$ in Fig. \ref{fig:0-pi3} (b). The $\theta _m$ pendulum has a long stay around $\theta _m = 2n\pi$, which can be understood by Fig. \ref{fig:0-pi3} (b).

Thus, we can obtain the transition from the $0$--$\pi$ phase mode to the $0$--$running$ phase mode by changing $m (0)$. 

 \begin{figure}[t]
\begin{center}
\includegraphics[width=0.9\linewidth]{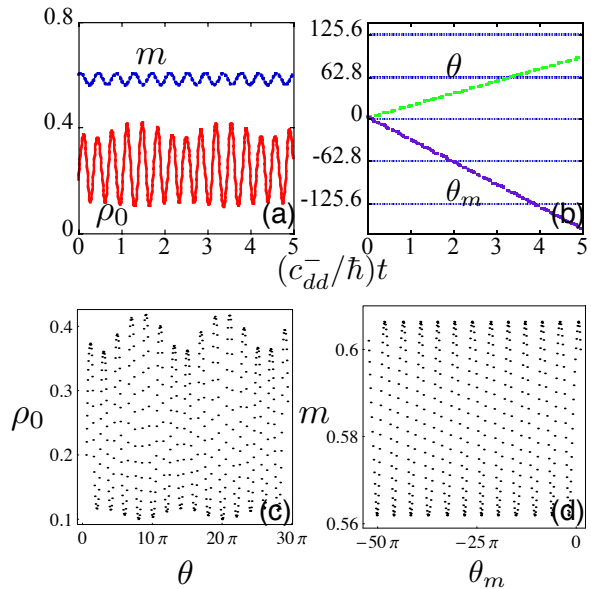} 
\caption{ (color online)  $running$--$running$ phase mode for $\rho _0 (0) = 0.6$, $m (0) = 0.2$, $\theta (0) = \pi/2$ and $\theta _m (0) = \pi$.}
\label{fig:run-run}
\end{center}
\end{figure}

Third, the $running$--$running$ phase mode is shown in Fig. \ref{fig:run-run}. The solutions show that the motions of the pendulums are rotations with different angular frequencies. 

 \begin{figure}[t]
\begin{center}
\includegraphics[width=0.9\linewidth]{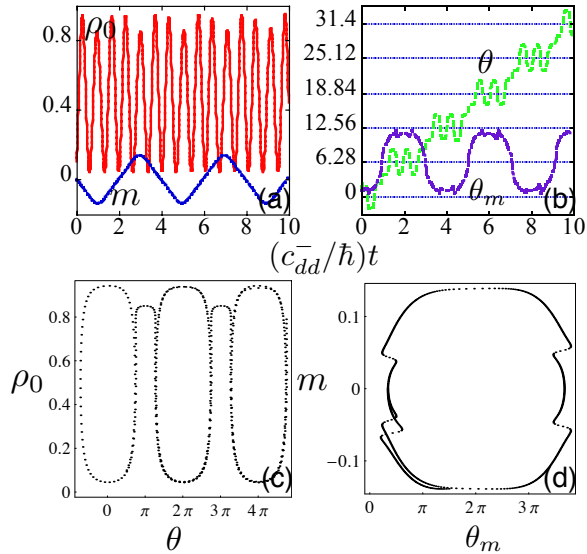} 
\caption{ (color online)  $2n\pi \ \& \ running$--$2\pi$ phase mode for $\rho _0 (0) = 0.1$, $m (0) = 0$, $\theta (0) = \pi/2$, $\theta _m (0) = \pi/3$.}
\label{fig:2npirun-2pi}
\end{center}
\end{figure}

Fourth, we present the $2n\pi \ \& \ running$--$2\pi$ phase mode in Fig. \ref{fig:2npirun-2pi}, where $n$ is an integer. In this mode, the dynamics of the two pendulums show strange and interesting motions. Especially, the $\theta$ pendulum repeats an oscillation around $\theta = 2n\pi$ and a rotation from $\theta = 2n\pi$ to $(2n+1)\pi$. The dynamics indicates a transition from the contour lines of the $2\pi$ phase mode to another line of the {\it running } phase mode, which is shown clearly in Fig. \ref{fig:2npirun-2pi} (c). Due to this transition, transitions should occur for $\dot{{\cal H}} = 0$  in $\{m, \theta _m\}$. Also, the $\theta _m$ pendulum has a time average of $\langle \theta \rangle = 2\pi$. However, the pendulum repeats a rotation from $\theta \simeq 0$ to $4\pi$ and returns to $\theta \simeq 0$.

 \begin{figure}[t]
\begin{center}
\includegraphics[width=0.9\linewidth]{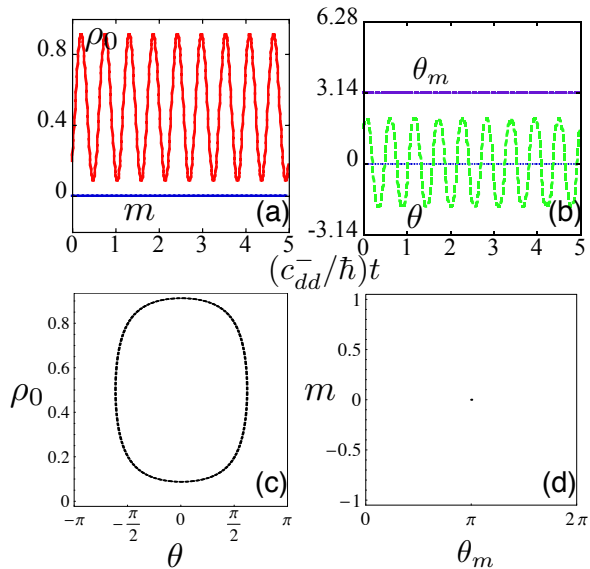} 
\caption{ (color online)  $single \ nonrigid \ pendulum$ phase mode for $\rho _0 (0) = 0.2$, $m (0) = 0$, $\theta (0) = \pi/2$, $\theta _m (0) = \pi$.}
\label{fig:single}
\end{center}
\end{figure}

 \begin{figure}[t]
\begin{center}
\includegraphics[width=0.9\linewidth]{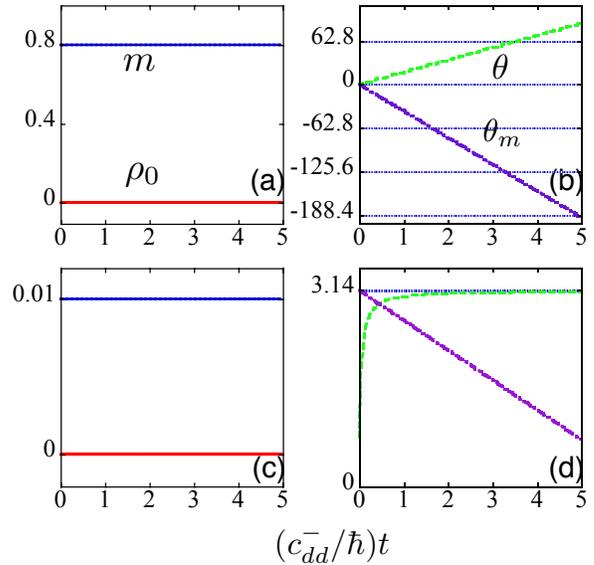} 
\caption{ (color online) The motion of two rigid pendulums for the initial conditions $\rho _0 (0) = 0$, $m (0) = 0.8$, $\theta (0) = 0$ and $\theta _m (0) = 0$ for (a) and (b), and $\rho _0 = 0$, $m =0.01$, $\theta = \pi/4$, and $\theta _m =\pi$ for (c) and (d).}
\label{fig:rigid}
\end{center}
\end{figure}

We obtain the motions of a single nonrigid pendulum in Fig. \ref{fig:single} and two rigid pendulum in Fig. \ref{fig:rigid}, namely the $single \ nonrigid \ pendulum$ and $two \ rigid \ pendulums$ phase modes. For the single pendulum, the $\theta$ pendulum oscillates around $\theta = 0$, whereas the length of the $\theta _m$ pendulum grows and shrinks without an oscillation because $\theta _m = 0$. The phase space in Fig. \ref{fig:single} (c) shows a single trajectory, indicating dynamics like that of a rigid pendulum. On the other hand, Fig. \ref{fig:rigid} (a) and (b) show the {\it running}--{\it running} mode for rigid pendulums given by the constants $\rho _0$ and $m$. Then, Fig. \ref{fig:rigid} (c) and (d) are very difficult solutions for understanding the dynamics. The $\theta$ pendulum exponentially deviates from the initial angle to $\theta = \pi$. The $\theta _m$ pendulum, however, exhibits the rotation of a rigid pendulum. 
	
\subsubsection{Contour lines of the Hamiltonian}\label{sssec:hamiltonian}

Here we discuss the various modes for two pendulums from the Hamiltonians.

It is impossible to plot the contour lines for the total Hamiltonian, namely Eq. (\ref{eq:H}), because of the four dimensions given by $\rho _0$, $m$, $\theta$ and $\theta _m$. By considering the Hamiltonian of the $\theta$ pendulum ${\cal H}_\theta$ and the $\theta _m$ pendulum ${\cal H} _{\theta _m}$, however, the features of the motions of the pendulums can be obtained.

\begin{figure}[t]
\begin{center}
\includegraphics[width=0.99\linewidth]{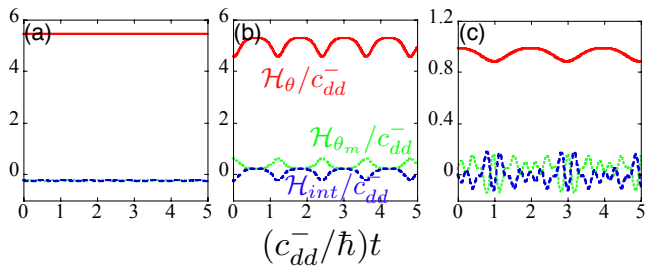} 
\caption{ (color online) Time development of the Hamiltonians. (a), (b), and (c) have the same initial conditions as for Fig. \ref{fig:0-pi1}, Fig. \ref{fig:0-pi2}, and Fig. \ref{fig:2npirun-2pi}, respectively. }
\label{fig:hamil}
\end{center}
\end{figure}

\begin{figure}[t]
\begin{center}
\includegraphics[width=0.99\linewidth]{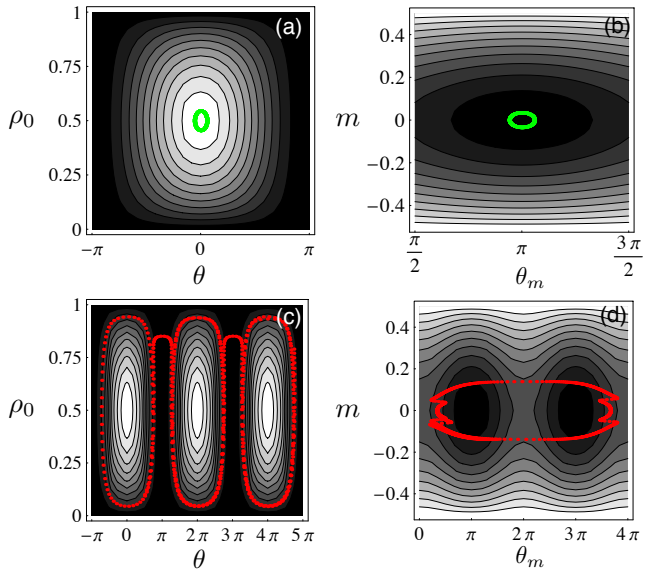} 
\caption{(color online) Trajectories and contour lines for the $0$--$\pi$, (a) and (b), and $2n\pi \   \& \ running$--$2\pi$ phase modes, (c) and (d). The trajectories are shown as green dots in the $0$--$\pi$ mode and red dots in the $2n\pi \   \& \ running$--$2\pi$ mode.  The lines are given by Eq. (\ref{eq:H_theta}) for $m = \langle m \rangle \simeq 0$ in (a) and (c) and Eq. (\ref{eq:H_theta_m}) for $\rho _0 = \langle \rho _0 \rangle \simeq 0.5$ in (b) and $0.47$ in (d). The magnitude of the Hamiltonians increases from black to white.}
\label{fig:contra}
\end{center}
\end{figure}

As mentioned in Sec. \ref{ss:jte}, ${\cal H}$ is a conserved value. However, ${\cal H}_\theta$, ${\cal H}_{\theta _m}$ and ${\cal H}_{int}$ are not conserved. Figure \ref{fig:hamil} shows the time development of these Hamiltonians, indicating the non-conservation, noting though that the values in (a) exhibit only small oscillations that do suggest energy conservation. Therefore, the trajectory in the phase spaces does not have to follow a contour line. As examples, we show the trajectories and contour lines for the $0$--$\pi$ and $2n\pi \ \& \ running$--$2\pi$ phase modes in Fig. \ref{fig:contra}. Naturally, the contour lines change with time because ${\cal H}_\theta$ and ${\cal H}_{\theta _m}$ include the parameters $m(t)$ and $\rho _0(t)$, which are time dependent. In Fig. \ref{fig:contra}, we plot the lines for the time averages $m = \langle m \rangle$ and $\rho _0 = \langle \rho _0 \rangle$ in ${\cal H}_{\theta }$ and ${\cal H}_{\theta _m}$. The trajectories of the $0$--$\pi$ phase mode almost follow the contour lines in Fig. \ref{fig:contra} (a) and (b). On the other hand, for the $2n\pi \ \& \ running$--$2\pi$ phase mode of (c) and (d), the trajectories do not follow the lines. However, the transition from the $0$ to $running$ phase modes is clearly seen in $\{\rho _0, \theta\}$. An important result is that the dynamics of the $\theta _m$ pendulum cannot be discussed as transitions between the $0$, $\pi$ and $running$ phase modes.

\section{CONCLUSION}\label{sec:con}

By introducing the single mode approximation into the spin-1 GP equations with the magnetic dipole--dipole interactions, we show the transitions between the three states, which are analogous with Josephson junctions of a ring of three superconductors. Furthermore, deriving the canonical equations of motion from the Josephson equations, the Josephson junctions are found to also be analogous with two nonrigid pendulums having interactions with each other. First, in order to consider a simple nonrigid pendulum, we assume that the shapes of the condensates are spheres. In this way, we review well known three phase nodes, the $0$, $\pi$ and $running$ phase modes. Second, we numerically solve the equations for two nonrigid pendulums, showing several motions: the $0$--$\pi$, $0$--$running$, $running$--$running$, $2n\pi \ \& \ running$--$ 2\pi$, $single \ pendulum$, and $two \ rigid \ pendulum$ phase modes. Finally, we discuss the transition between the modes from the non-conserved Hamiltonians ${\cal H}_\theta$, ${\cal H}_{\theta _m}$, and ${\cal H}_{int}$. 

We consider that the Josephson effect, which has been found in various phenomena from condensed matter physics to classical physics, is a universal and important physical phenomenon. In this study, the effect was discussed only in relation to BECs. However, we expect that the study of the Josephson effect of the three states will be useful in many other fields beyond atomic BECs.

\section{acknowlegement}\label{sec:ack}

M. Y. acknowledges the support of a Research Fellowship of the Japan Society for the Promotion of Science for Young Scientists (Grant No. 209928). M. T. acknowledges the support of a Grant-in Aid for Scientific Research from JSPS (Grant No. 21340104) and a Grant-in-Aid for Scientific Research on Priority Areas from MEXT (Grant No. 17071008).

\end{document}